\documentclass[twocolumn,showpacs,floats,floatfix,superscriptaddress,aps,pra]{revtex4}
\usepackage{amsfonts}
\usepackage{amssymb}
\usepackage{amsmath}
\usepackage{graphicx}
\usepackage{bm}
\usepackage{color}
\usepackage{epsfig}
\usepackage{ifthen}
\usepackage{color}

\begin{document}

\author{A. A. Rangelov}
\affiliation{Department of Physics, Sofia University, James Bourchier 5 blvd., 1164
Sofia, Bulgaria}
\author{N. Karchev}
\affiliation{Department of Physics, Sofia University, James Bourchier 5 blvd., 1164
Sofia, Bulgaria}
\title{Coulomb interaction revised in the presence of material with negative
permittivity}
\date{\today }

\begin{abstract}
Coulomb law is one of the fundamental laws in Physics. It describes the
magnitude of the electrostatic force between two electric charges.
Counterintuitively the repulsion force between two equal electric charges in
a vacuum, stated by the Coulomb law, turn into the attraction force between
the same electric charges when they are placed next to a material with
negative permittivity and the distance between them is larger than some
critical distance. As a result the equally charged particles ``crystallize"
occupying equilibrium positions. We prove this claim with the method of
images for two charged particles placed next to a material with negative
permittivity.
\end{abstract}

\maketitle



\section{Introduction \label{Sec-Intro}}


Most materials have positive permittivity. If such a material is placed in
an electric field, then the direction of the field induced inside the
material will have the same orientation as the applied field. Whereas the
field inside a material with negative permittivity would be oriented in the
opposite direction to the applied field. The most popular substances with
negative permittivity are gaseous plasmas and solid-state plasmas \cite%
{Veselago,Skobov}. In a plasma with no magnetic field, the permittivity $%
\varepsilon /\varepsilon_{0}$ ($\varepsilon_{0}$ -vacuum permittivity) is
given by%
\begin{equation}  \label{omega}
\varepsilon /\varepsilon_{0}=1-\omega_{0}^{2}/\omega ^{2},
\end{equation}
where $\omega_{0}=\sqrt{4\pi Ne^{2}/m}$ is the plasma frequency, $N$ is the
concentration of the carriers, $e$ is their charge, and $m$ is their mass.
It is not hard to check that when the frequency $\omega $ is smaller
compared to the plasma frequency $\omega _{0}$, then $\varepsilon $ is
negative.

It has been shown that wire structures with lattice spacing of the order of
a few millimeters behave like a plasma with a resonant frequency, $%
\omega_{0} $, in the GHz regions. These metamaterials gain their properties
from the structure rather than the composition \cite{Sievenpiper,Pendry}.
Such materials attract growing interest for the last decade from theoretical
and experimental perspectives. The presence of simultaneously negative
permeability and permittivity in these materials \cite{Pendry,Shelby} lead
to many unusual effects which find various applications in areas such as
perfect and hyper lenses \cite%
{Pendry2,Melville,Fang,Taubner,Jacob,Liu,Smolyaninov}, suppression of
spontaneous emission of an atom in front of a mirror made by metamaterial
\cite{Kastel}, magnification of objects that are smaller than the wavelength
\cite{Wee} and creation of second-harmonic generation \cite{Klein}.

Veselago made a theoretical study of materials with negative electrical
permittivity, $\varepsilon $ and negative magnetic permeability, $\mu $ in
his seminal paper \cite{Veselago}. The conclusion of his article is that
this type of materials support electromagnetic waves description, but the
energy flow, directed by the Poynting vector, is in the opposite direction
to the wave vector. This means that rays travel in the opposite direction to
waves. The most important result of Vaselago's work is that when both $%
\varepsilon $ and $\mu $ are negative, the refraction index, defined by the
equation $n^{2}=\varepsilon \mu $, is negative
\begin{equation}
n\,=\,-\sqrt{\varepsilon \mu }.
\end{equation}%
The key experimental consequence is the rather unusual manifestation of
Snell's law. In the passage of a ray of light from one medium with positive
refraction index $n_{1}>0$ into another, with $n_{2}<0$ the Snell's law is
satisfied but $n_{1}/n_{2}$ is negative. In materials with negative
refraction index the Cherenkov effect is reversed, just like the Doppler
effect \cite{Veselago2}.

The Veselago theory is focused on the optical properties of materials with
negative permittivity and permeability. In the present paper we consider the
electric forces in the presence of material with negative permittivity. We
explore theoretically the force acting on a single charged particle and the
interaction between two charged particles placed next to a material with
negative permittivity. Our examination relies on the Pendry's criterion for
validity of electrostatic limit in the present problem \cite{Pendry4}: the
wavelength, corresponding to the frequency $\omega$ in Eq.\ref{omega},
should be longer compare with the distance of the charged particles to the
material with negative permittivity. We show, using the method of images
\cite{Greiner,Griffiths,Jackson}, that for negative permittivity of the
material, the force between the material and the charged particle can be
attractive or repulsive and can even have a bigger value compared to the
conventional materials. Furthermore, the repulsion force between two equally
charged electric particles in a vacuum, stated by the Coulomb law, turn into
the attraction force between the same electric charges when they are placed
next to a material with negative permittivity and the distance between them
is larger than some critical distance. As a result the two equally charged
particles ``crystallize" occupying equilibrium states.


\section{Method of images for material with negative permittivity\label%
{Sec-Method of images}}


\subsection{A point charge above dielectric plane \label{subsec-dielectric
plane}}


Initially we start our examination with the simplest possible case of a
single point charge $q$, which is embedded in a semi-infinite dielectric
with permittivity $\varepsilon _{1}$ at a distance $d$ away from a plane
interface that separates the first medium from another semi-infinite
dielectric with permittivity $\varepsilon _{2}$ (Fig.\ref{single point
charge}). We want to find what the force acting on the charge $q$ is. From
the point of view of Mathematics our problem is to solve Poisson's equation
in a region with permittivity $\varepsilon _{1}$, and a point charge $q$,
subject to the boundary conditions at the plane interface that separates the
first medium from the second. The solution of this problem is easily found
by the method of images \cite{Greiner,Griffiths,Jackson}. This method is a
powerful and easy way to handle solutions of differential equations, in
which the domain of the sought function is extended by the addition of its
mirror image with respect to a symmetry hyperplane. In our case the
electrical potential will be fully reconstructed in the space occupied by
the dielectric $\varepsilon _{1}$ if we place an image of $q$ with a charge
magnitude%
\begin{equation}
q^{\prime }=\frac{\varepsilon _{1}-\varepsilon _{2}}{\varepsilon
_{1}+\varepsilon _{2}}q
\end{equation}%
at a distance $d$ away from a plane in the space occupied by the dielectric $%
\varepsilon _{2}$ \cite{Greiner,Griffiths,Jackson} (Fig.\ref{single point
charge}). Therefore the net force acting on the charge $q$ is
\begin{equation}
F=\frac{1}{4\pi \varepsilon _{1}}\frac{qq^{\prime }}{\left( 2d\right) ^{2}}=%
\frac{q^{2}}{16\pi \varepsilon _{1}d^{2}}\frac{\varepsilon _{1}-\varepsilon
_{2}}{\varepsilon _{1}+\varepsilon _{2}},  \label{net
electrical force}
\end{equation}%
which force could be attractive or repulsive depending on the ratio $\left(
\varepsilon _{1}-\varepsilon _{2}\right) /\left( \varepsilon
_{1}+\varepsilon _{2}\right) $. The maximal attraction force for
conventional materials ($\varepsilon _{1}>0$, \thinspace \thinspace\ $%
\varepsilon _{2}>0$) is \cite{Greiner,Griffiths,Jackson}
\begin{equation}
F=-\frac{q^{2}}{16\pi \varepsilon _{0}d^{2}}.  \label{conductor force}
\end{equation}%
We are interested in the case when the first media is a vacuum, but the
second media is a material with negative permittivity ($\varepsilon
_{1}=\varepsilon _{0},\,\,\varepsilon _{2}<0$). In this case the net force
acting on the charge $q$ is%
\begin{equation}
F=\frac{q^{2}}{16\pi \varepsilon _{0}d^{2}}\frac{\varepsilon
_{0}-\varepsilon _{2}}{\varepsilon _{0}+\varepsilon _{2}}.
\label{metamaterial electrical force}
\end{equation}%
If the permittivity $\varepsilon _{2}$ of the material satisfies
\begin{equation}
\left\vert \varepsilon _{2}\right\vert <\varepsilon _{0},
\end{equation}
then the force from Eq.(\ref{metamaterial electrical force}) has a positive
sign, therefore the charge is repelled from this material. The force from
Eq.(\ref{metamaterial electrical force}) is attractive if
\begin{equation}
\left\vert \varepsilon _{2}\right\vert >\varepsilon _{0}
\end{equation}

Obvious singularity happens in Eq.(\ref{metamaterial electrical force}) when
$\varepsilon _{2}=-\varepsilon _{0}$. One might make a mistake thinking that
this singularity leads to an infinite force, but one should note that
dispersionless material is an abstraction and a finite positive imaginary
component of the permittivity always exists \cite{Pendry4}. The imaginary
component of the permittivity can no longer be neglected when the difference
between the real parts of the two dielectric permittivities is a small
number, which resolve the paradox with the infinite force.

\begin{figure}[tb]
\centerline{\epsfig{width=75mm,file=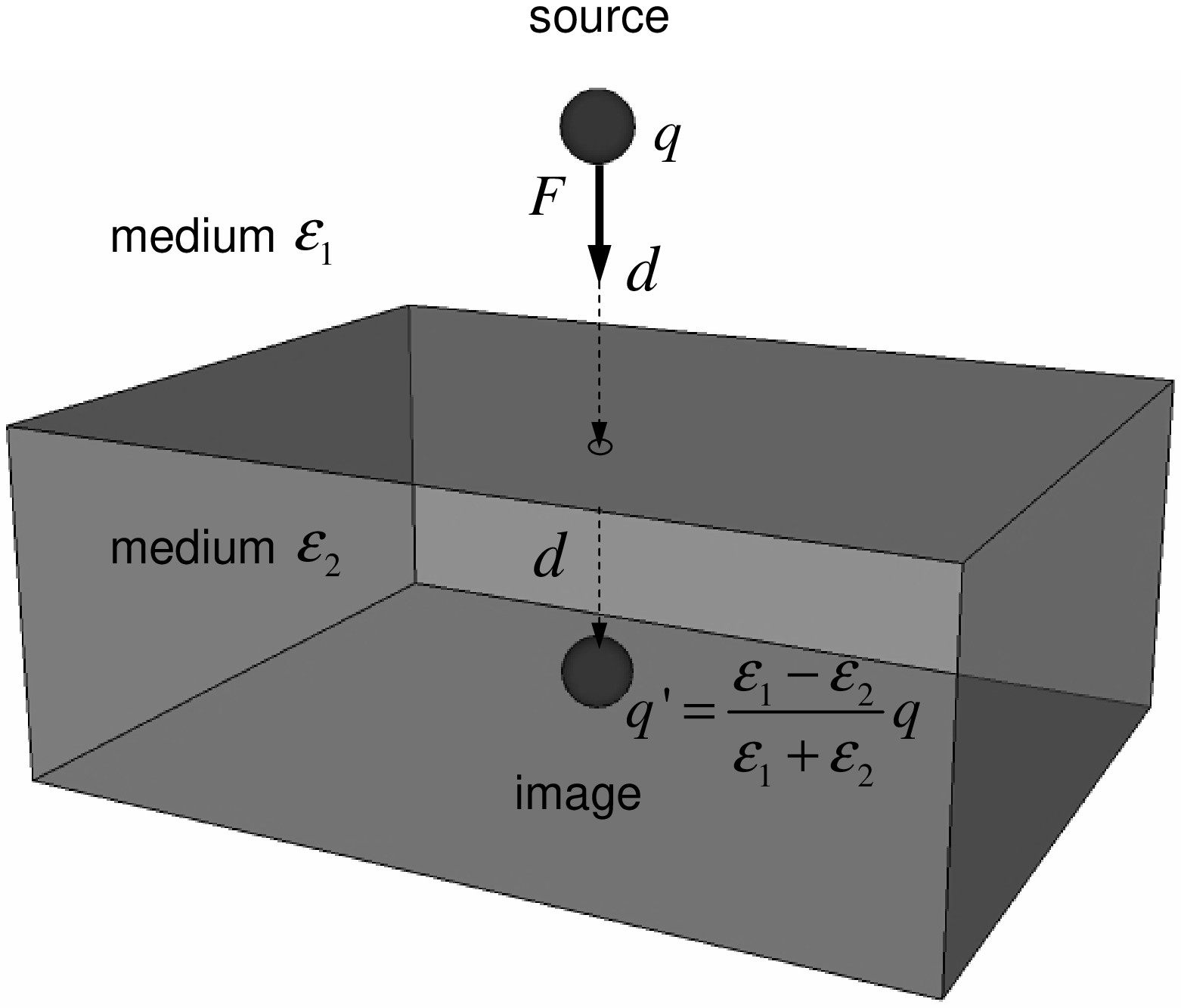}}
\caption{A point charge $q$ embedded in a semi-infinite dielectric with
permittivity $\protect\varepsilon _{1}$ at a distance $d$ away from a plane
interface that separates the two mediums and its image charged symmetrically
placed in a semi-infinite dielectric with permittivity $\protect\varepsilon %
_{2}$.}
\label{single point charge}
\end{figure}

An important case is when we can neglect the imaginary part of the
permittivity $\varepsilon _{2}$. Then the force from Eq.(\ref{metamaterial
electrical force}) could be repulsive and bigger compared to the module of
the force from Eq.(\ref{conductor force}). For example if $\varepsilon
_{2}=-\varepsilon _{0}/2$, then the net force acting on the charge $q$ is
\begin{equation}
F=\frac{3q^{2}}{16\pi \varepsilon _{0}d^{2}},  \label{strong repulsive force}
\end{equation}%
which is a repulsive force and three times stronger than the module of the
force that the same charge $q$, placed at the same distance $d$, feels from
the plane interface of a metal (see Eq.(\ref{conductor force})).

\subsection{Two point charges above a dielectric plane\label{subsec-two
point charges}}

\begin{figure}[htb]
\centerline{\epsfig{width=75mm,file=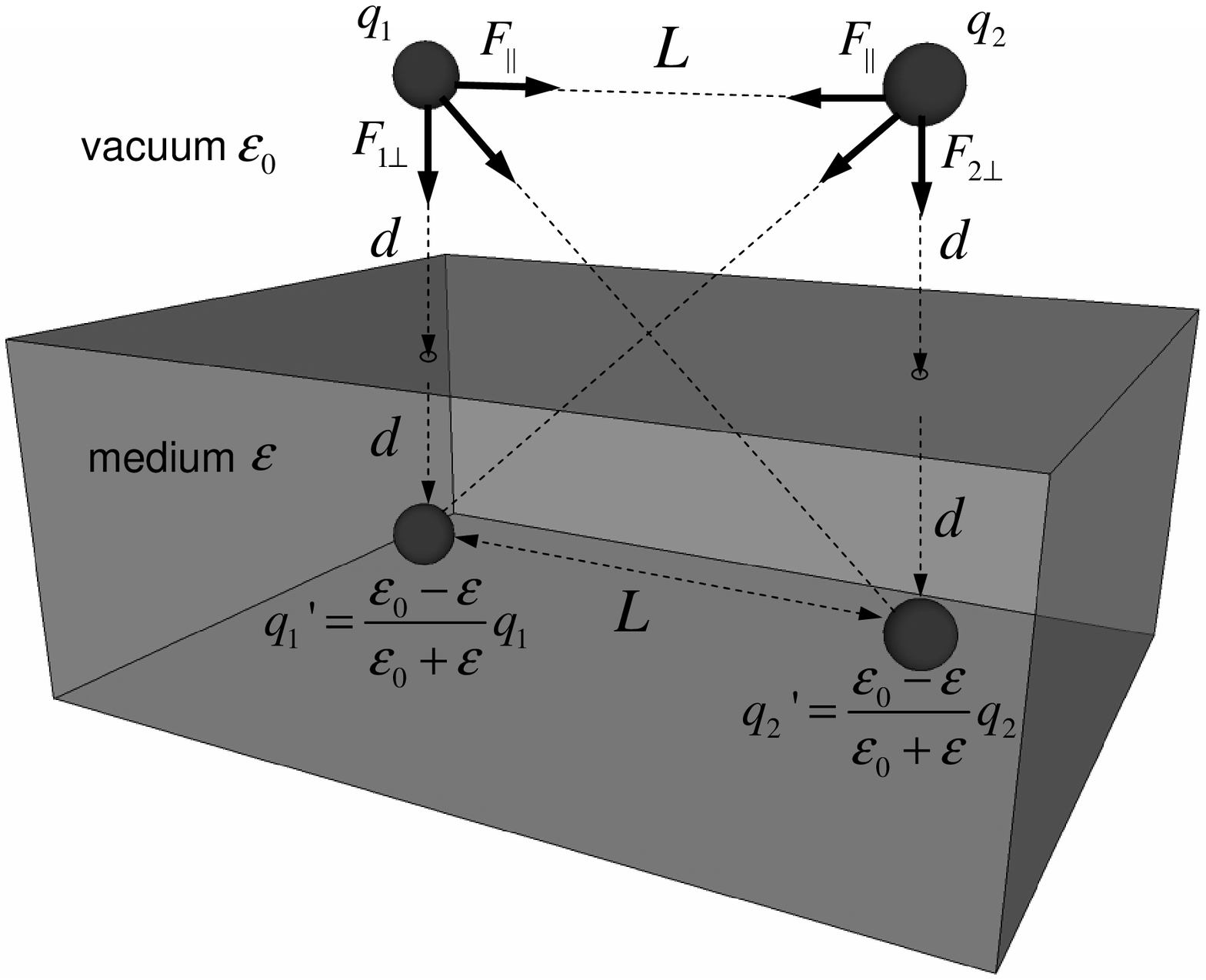}}
\caption{Two point charges $q_{1}$ and $q_{2}$ placed in a vacuum at a
distance $d$ away from a plane interface that separates the vacuum with a
semi-infinite dielectric with permittivity $\protect\varepsilon $ and their
images $q_{1}^{\prime }$ and $q_{2}^{\prime }$.}
\label{two point charges}
\end{figure}

Even a more interesting situation with a counterintuitive result is the case
when two charged particles are placed in a vacuum next to a material with
negative permittivity $\varepsilon $ (Fig.\ref{two point charges}). To
examine the situation in more details, we consider the case when the two
charged particles $q_{1}$ and $q_{2}$ are in a vacuum at a distance $d$ away
from the material with negative permittivity $\varepsilon $. The method of
images states that the system is described equivalently by two images for
each charge $q_{1}$ and $q_{2}$ placed at a distance $d$ away from the
surface of the material $\varepsilon $ (Fig.\ref{two point charges}) with
charges
\begin{eqnarray}
q_{1}^{\prime } &=&\frac{\varepsilon _{0}-\varepsilon }{\varepsilon
_{0}+\varepsilon }q_{1}, \\
q_{2}^{\prime } &=&\frac{\varepsilon _{0}-\varepsilon }{\varepsilon
_{0}+\varepsilon }q_{2}.
\end{eqnarray}
For convenience let us represent the force in two components:
\begin{equation}
F_{\parallel }=\frac{q_{1}q_{2}}{4\pi \varepsilon _{0}L^{2}}\left( 1+\frac{%
\varepsilon _{0}-\varepsilon }{\varepsilon _{0}+\varepsilon }\frac{L^{3}}{%
\left( L^{2}+4d^{2}\right) ^{3/2}}\right) ,  \label{parallel
component}
\end{equation}%
\begin{equation}
F_{\alpha \perp }=\frac{q_{\alpha }^{2}}{16\pi \varepsilon _{0}d^{2}}\frac{%
\varepsilon _{0}-\varepsilon }{\varepsilon _{0}+\varepsilon }\left( 1+\frac{%
q_{1}q_{2}}{q_{\alpha }^{2}}\frac{8d^{3}}{\left( L^{2}+4d^{2}\right) ^{3/2}}%
\right) ,  \label{perpendicular component}
\end{equation}%
where $\alpha =1$ or $2$ and $L$ is the distance between the charged
particles. The perpendicular to the material component of the force can be
thought of as an effective force between the charged particles and the
material, while the parallel component as an effective interaction between
the particles.

The expression in the brackets in Eq.(\ref{parallel component}) is positive
when $\varepsilon >0$ for arbitrary values of $d$ and $L$. This means that
the character of the interaction (attractive or repulsive) is the same as
that stated by the Coulomb low.

When $\varepsilon <0$ and $|\varepsilon |>\varepsilon _{0}$ there is a
critical distance between the particles $L_{cr}$ at which the interaction is
zero $(F_{\parallel }(L_{cr})=0)$
\begin{equation}
L_{cr}=\frac{2d}{\sqrt{\left( \frac{|\varepsilon |+\varepsilon _{0}}{%
|\varepsilon |-\varepsilon _{0}}\right) ^{2/3}-1}}.
\end{equation}%
From Eq.(\ref{parallel component}) it follows that: when the two particles
are closer $(L<L_{cr})$, the second term in the brackets is smaller than the
first one and the interaction between the charges is the same as stated by
the Coulomb law. But when the distance between the particles is larger than
the critical one $(L_{cr}<L)$, the expression in the brackets of Eq.(\ref%
{parallel component}) is negative and the force has an opposite sign
compared to the force given by the Coulomb law.

\begin{figure}[tb]
\centerline{\epsfig{width=75mm,file=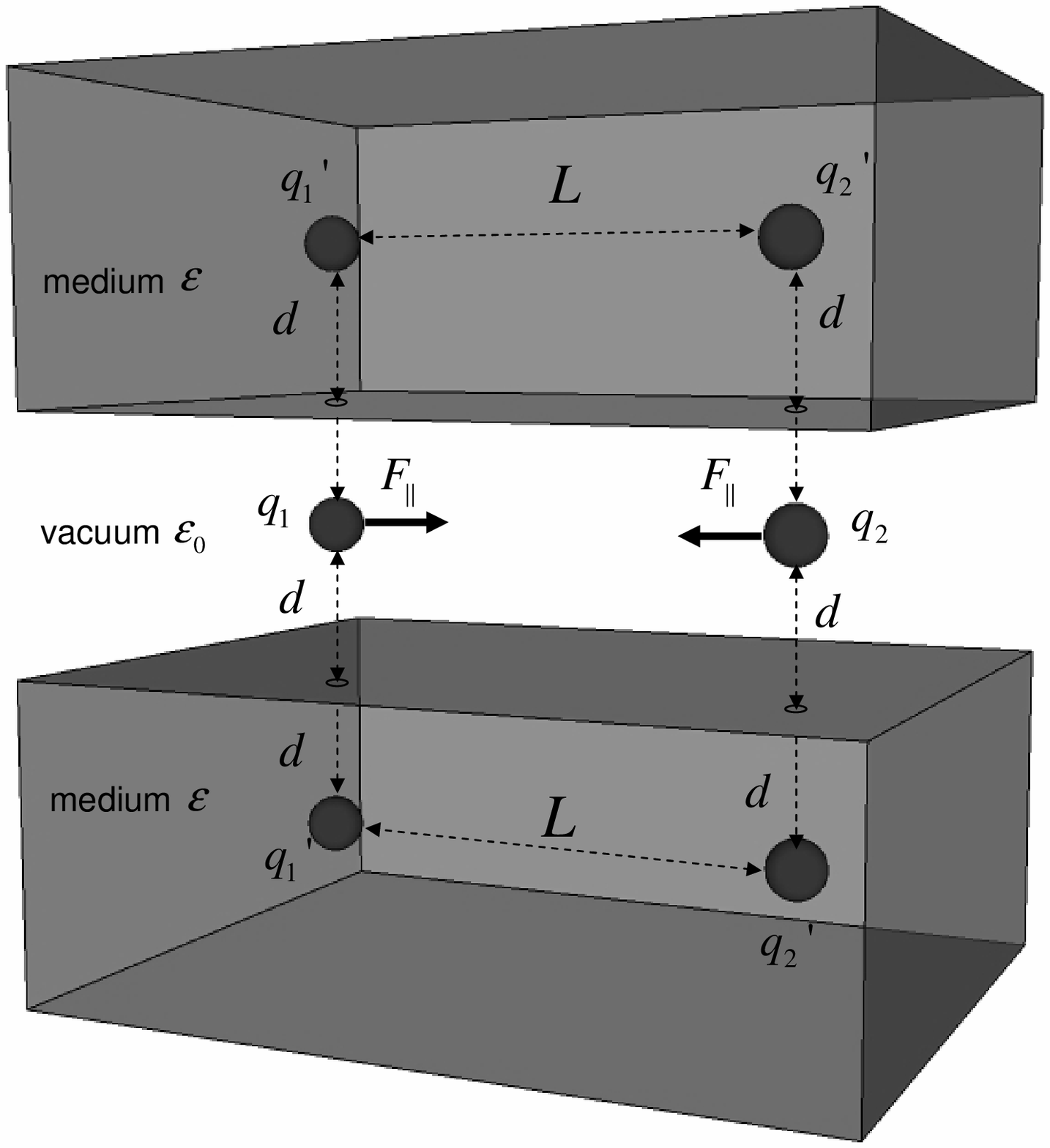}}
\caption{The symmetric geometry case of two charged particles $q_{1}$ and $%
q_{2}$, placed in a middle way between two identical semi-infinite
metamaterials with permittivity $\protect\varepsilon $ and their images $%
q_{1}^{\prime }$ and $q_{2}^{\prime }$.}
\label{the symmetric geometry}
\end{figure}

The total perpendicular component of the force can be zero $(L_{\alpha
\perp}=0)$ if a symmetric geometry is realized. For example this is the case
when the charged particles are placed in a vacuum sandwich with two
identical semi-infinite materials with negative permittivity $\varepsilon $
(Fig.\ref{the symmetric geometry}). In the particular case when the two
particles have equal charges $q_1=q_2=q$ and $|\varepsilon|>\varepsilon_0$,
the interaction between them is
\begin{equation}
F_{\parallel}=\frac{q^2}{4\pi \varepsilon _{0}L^{2}}\left( 1-2 \frac{%
|\varepsilon |+\varepsilon_{0} }{|\varepsilon |-\varepsilon_{0} }\frac{L^{3}%
}{\left( L^{2}+4d^{2}\right) ^{3/2}}\right) ,  \label{parallel
component2}
\end{equation}%
The factor $2$ in the brackets in Eq (\ref{parallel component2}) is owing to
the geometry of the system. It changes when the geometry of the system is
altered. The critical distance at which the force between particles Eqs (\ref%
{parallel component2}) is zero $(F_{\parallel}(L^{\prime }_{cr})=0)$ depends
on the geometry too
\begin{equation}
L^{\prime }_{cr}=\frac{2d}{\sqrt{\left(2 \frac{|\varepsilon |+\varepsilon
_{0}} {|\varepsilon |-\varepsilon _{0}}\right) ^{2/3}-1}} .
\label{criticalL}
\end{equation}

The force, Eq (\ref{parallel component2}), between two equal charged
particles is repulsive when they are closer than the critical distance $%
L^{\prime }_{cr}$, and it is attractive when the distance is larger than the
critical one. As a result the particles ``crystallize" occupying equilibrium
positions at a distance equal to the critical distance $L^{\prime }_{cr}$.

The scheme could be extended to many particles. By increasing the number of
the particles, the critical distance decreases. In this way we can make an
array of equally charged particles situated at a very small distance from
each other. This permits us to construct a new type of equally charged
particles trap.

By making the permittivity $\varepsilon$ to vary from negative to positive,
we release the particles and they will have a kinetic energy depending on
the critical distance between them. This enables us to accelerate the
charges in a new fashion.

\section{Conclusions\label{Sec-Conclusions}}

The examined force between a charged particle and a material with negative
permittivity has the potential to be not only a curious and intriguing
example of what the artificial material can do, but also a useful technique
to levitate small objects with charges and therefore to make frictionless
devices.
\acknowledgments
This work was supported by the European Commission projects EMALI and
FASTQUAST, and the Bulgarian NSF grants VU-F-205/06, VU-I-301/07,
DO02-90/08, DO02-264/08 and Sofia University grants 020/2009, 095/2009.

\end{document}